\documentclass[showpacs,twocolumn,pre]{revtex4}
\usepackage[dvips]{epsfig}
\newcommand{\clV}{\mathcal{ V}}

\newcommand{\clM}{\mathcal{ M}}

\newcommand{\clC}{{\cal C}}
\newcommand{\clL}{{\cal L}}

\newcommand{\hclL}{\hat{\cal L}}

\newcommand{\clE}{\mathcal{E}}

\newcommand{\tlG}{\tilde{G}}
\newcommand{\rgl}{\rangle}
\newcommand{\lgl}{\langle}
\newcommand{\ep}{\epsilon}
\newcommand{\vep}{\varepsilon}
\newcommand{\be}{\begin{equation}}
\newcommand{\ee}{\end{equation}}
\newcommand{\bea}{\begin{eqnarray}}
\newcommand{\eea}{\end{eqnarray}}

\begin{document}

\title{Geometrical enhancement of the electric field:  \\
Application of fractional calculus in nanoplasmonics}

\author{Emmanuel Baskin and Alexander Iomin}

\affiliation{Department of Physics and Solid State Institute,
Technion, Haifa, 32000, Israel}

\begin{abstract}
 We developed an analytical approach, for a wave propagation in
metal-dielectric nanostructures in the quasi-static limit. This
consideration establishes a link between fractional geometry of
the nanostructure and fractional integro-differentiation. The
method is based on fractional calculus and permits to obtain
analytical expressions for the electric field enhancement.
\end{abstract}


\maketitle

\section{Introduction}
An interplay between the light and fractional nanostructures leads
to a local giant enhancement of the electric field. This is one of
the main effect of \textit{nanoplasmonics} \cite{Shalaev,Stockman}
that concerns with the electromagnetic field on a nanoscale less
then the wavelength inside a composite matter. There is a vast
literature starting from seminal results on optics in fractal
smoke \cite{berrypercival} with further developing, where strong
fluctuations of a local electromagnetic field are taken into
account, see detailed discussion in \cite{Markel} and recent
reviews \cite{Shalaev,Stockman,stRev}. A suitable mechanism of
these phenomena have been proposed in many studies
\cite{Shalaev,stRev,st3,st4}), and it relates to the optical
properties of metal nanoparticles, and, mainly, is due to their
ability to produce giant and highly localized electromagnetic
fields inside composite nanostructure. We are speaking about a
\textit{localized} surface plasmon resonance (SPR) (see, for
example Ref. \cite{Shalaev,st3}). Localized surface plasmons are
charge density oscillations confined to conducting (metallic)
nanostructures. A strong enhancement of the electromagnetic field
is possible if its frequency is near the frequency of these
oscillations. An analogous phenomenon,  also known in optics, is a
so-called enhanced Raman scattering. This remarkable phenomenon
has been firstly observed more than thirty years ago,  see
\textit{e.g.}, \cite{stRev,st4} and properties of the Raman
spectroscopy, has been used for a variety of applications.
Theoretical explanation follow these experiments \cite{st4} and
still attract much attention \cite{Shalaev,Stockman,st3,st4}.

In this paper we suggest an analytical consideration, based on
fractional calculus, purposely to obtain an analytical expression
for the electric field enhancement in nanoplasmonics. This
approach establishes a relation between fractal geometry of the
nanostructure and fractional integro-differentiation. It is based
on averaging an extensive physical value expressed by means of a
smooth function over a fractal set that leads to fractional
integration \cite{bi2011}. We suggest a coarse graining procedure
of the electric field in the Maxwell equation to treat a charge
density term, which diverges everywhere in the case of fractional
metal-dielectric composite media. This smoothing procedure makes
possibility to obtain an equation for an electric field in a
closed form. Our approach is based on a seminal result
\cite{NIG92}, where this scheme has been suggested for a Cantor
set. In its eventual form, it has been presented in Ref.
\cite{NIG98} (in Ch. 5). The main idea of embedding or filtering
matter inside a fractal is the construction of a convolution
integral
\be\label{fpe1}  %
M(x)=W(x)\star f(x)=\int_0^xW(x-y)f(y)dy\, ,  \ee  %
where the function $W(x)$ obeys the following scaling relation
$W(x)=\frac{1}{a}W(bx)$ with a solution $W(x)=t^{\alpha}A(x)$,
where $\alpha=\frac{\ln a}{\ln b}$ and $A(x)=A(bx)$ is a
log-periodic function with a period $\ln b$. When this scaling
corresponds to a Cantor set with $a=2$, and $A(x)$ is defined
explicitly, one performs averaging over this period \cite{NIG98}
and obtains
the convolution integral in the form of a fractional integral
\be\label{fpe3}   %
\lgl{M(x)}\rgl
=\frac{\clV(\alpha)}{\Gamma(\alpha)}\int_0(x-x')^{\alpha-1}f(x')dx'
\, ,  \ee %
where $\clV(\alpha)=\frac{2^{-1+\alpha/2}}{\ln 2}$ and
$\Gamma(\alpha)$ is a gamma function. One should recognize that
the coarse graining procedure due to the averaging over the period
$\ln b$ is important for the application of fractional calculus,
and ``a bond between fractal geometry and fractional calculus''
can be established on some coarse graining geometry \cite{NIG98}.
This mathematical construction is relevant to study of
electrostatics of real composite structures in the framework of
coarse-graining Maxwell's equation. The main objective of the
preset research is an \textit{analytical} derivation of the
electric field in a composite dielectric, which is subject to an
external high-frequency electric field at the conditions when the
quasistatic limit is valid.

\section{Fractional calculus briefly}
Extended reviews of fractional calculus can be found
\textit{e.g.}, in \cite{oldham,SKM,podlubny}. Fractional
integration of the order of $\alpha$ is defined by the operator
$${}_aI_x^{\alpha}f(x)=
\frac{1}{\Gamma(\alpha)}\int_a^xf(y)(x-y)^{\alpha-1}dy\, , $$
where $\alpha>0,~x>a$. Fractional derivation was developed as a
generalization of integer order derivatives and is defined as the
inverse operation to the fractional integral. Therefore, the
fractional derivative is defined as the inverse operator to
${}_aI_x^{\alpha}$, namely $
{}_aD_x^{\alpha}f(x)={}_aI_x^{-\alpha}f(x)$ and
${}_aI_x^{\alpha}={}_aD_x^{-\alpha}$. Its explicit form is
$${}_aD_x^{\alpha}f(x)=
\frac{1}{\Gamma(-\alpha)}\int_a^xf(y)(x-y)^{-1-\alpha}dy\, . $$
For arbitrary $\alpha>0$ this integral diverges, and, as a result
of this, a regularization procedure is introduced with two
alternative definitions of ${}_aD_x^{\alpha}$. For an integer $n$
defined as $n-1<\alpha<n$, one obtains the Riemann-Liouville
fractional derivative of the form
\begin{equation}\label{mt1a}   %
{}_a^{RL}D_x^{\alpha}f(x)=\frac{d^n}{dx^n}{}_aI_x^{n-\alpha}f(x)\,
,
\end{equation}
and fractional derivative in the Caputo form
\begin{equation}\label{mt1b}  %
{}_a^CD_x^{\alpha}f(x)= {}_aI_x^{n-\alpha}\frac{d^n}{dx^n}f(x)\, .
\end{equation}  %
There is no constraint on the lower limit $a$. For example, when
$a=0$, one has
${}_0^{RL}D_x^{\alpha}x^{\beta}=\frac{x^{\beta-\alpha}
\Gamma(\beta+1)}{\Gamma(\beta+1-\alpha)}\, , ~\alpha,\beta>0$.
This fractional derivation with the fixed low limit is also called
the left fractional derivative. Another important property is
$D^{\alpha}I^{\beta}=I^{\beta-\alpha}$, where other indexes are
omitted for brevity's sake. A convolution rule for the Laplace
transform for $0<\alpha<1$
\begin{equation}\label{mt3}
\clL[{}I_x^{\alpha}f(x)]=s^{-\alpha}\tilde{f}(s)
\end{equation}
is commonly used in fractional calculus as well.

\section{Quasi-static electromagnetic field}
We consider a sample of a size $R$ which consists of a fractal
metal nanosystem embedded in a dielectric host medium. The system
is subject to an external electromagnetic field $\mathbf{E}_0(t)$
at an optical frequency $\omega$, and we assume that the
fractal-host inhomogeneities are much smaller than the light
wavelength $\lambda=2\pi c/\omega$: $l_0\ll \lambda$, where $c$ is
the light speed and $l_0$ is a minimal size of the self-similarity
of the fractal volume. In this case a quasi-static approximation
is valid \cite{LLVIII}. Therefore, the nanosystem  contains a
metallic fractal with a fractal volume $V_D\sim R^D$ and the
permittivity $\vep_m(\omega)$, which depends on optical frequency
$\omega$, while the dielectric host has the volume $V_h$ and the
permittivity $\vep_d$. Following Refs.~\cite{st3,st4}, the wave
magnetic component is not considered, since the magnetic
permeability $\mu=1$ in both media. The electric component
satisfies the Maxwell equation in the Fourier frequency domain
\be\label{qsef1}   %
\nabla\cdot\Big[\vep(\mathbf{r},\omega)\mathbf{E}(\mathbf{r},\omega)\Big]=0\, . \ee   %
The permittivity can be expressed by means of a characteristic
function $\chi(\mathbf{r})$ \cite{st3,st4,bi2011} in the form
$\vep(\mathbf{r},\omega)=\vep_m(\omega)\chi(\mathbf{r})+\vep_d[1-\chi(\mathbf{r})]$,
where the characteristic function inside the fractal is
$\chi(\mathbf{r})=1$, $\mathbf{r}(x,y,z)\in V_D$, while inside the
dielectric host it reads $\chi(\mathbf{r})=0$,
$\mathbf{r}(x,y,z)\in V_h$. At the boundaries the electric field
consists of the external field and an induced field due to
polarization of the fractal nanostructure. Splitting the electric
field into the two components $
\mathbf{E}(\mathbf{r})=\tilde{\mathbf{E}}(\mathbf{r})+\mathbf{E}_1(\mathbf{r})$,
we looking for the reaction of the nanosystem on the external
field, where $\tilde{\mathbf{E}}(\mathbf{r})$ is the electric
field induced by $\mathbf{E}_0$ in the homogeneous nanosystem,
when $\chi(\mathbf{r})=0$ for $\forall ~ \mathbf{r}$, or
$\chi(\mathbf{r})=1$ for $\forall ~ \mathbf{r}$ and
$\nabla\tilde{\mathbf{E}}(\mathbf{r})=0$. Here
$\mathbf{E}_1(\mathbf{r})$ results from inhomogeneity of the
nanostructute. Thus Eq. (\ref{qsef1}) can be rewritten in the form
\bea\label{qsef1_a}  %
\chi(\mathbf{r})\nabla\cdot\mathbf{E}_1(\mathbf{r})&-&q(\omega)
\nabla\cdot\mathbf{E}_1(\mathbf{r})+
\mathbf{E}_1(\mathbf{r})\cdot\nabla\chi(\mathbf{r}) \nonumber  \\
&=&
-\tilde{\mathbf{E}}(\mathbf{r})\cdot\nabla\chi(\mathbf{r})\, , \eea   %
where $q(\omega)=\frac{\vep_d}{\vep_d-\vep_m(\omega)}$ is a
spectral parameter \cite{bergman}. The boundary condition for the
normal component of the dielectric shift is
$\mathbf{D}_n(\mathbf{r}\in S)=\mathbf{E}_{n}(\mathbf{r}\in S)$,
where $S$ means the sample boundaries.
%

The main complication of treating Eq. (\ref{qsef1_a}) is the
polarization charge density term $\sim \nabla\chi(\mathbf{r})$.
Since the characteristic function $\chi(\mathbf{r})$ is
discontinuous, $\nabla\chi(\mathbf{r})$ diverges everywhere.
Therefore, the next step of our consideration is coarse-graining
the electric field, which is averaging Maxwell's Eq.
(\ref{qsef1_a}). This procedure relates to integration with the
characteristic function
$\int\chi(\mathbf{r})\nabla\mathbf{E}_1(\mathbf{r})dV$ and
evaluation of the integral
$\int\nabla\chi(\mathbf{r})\cdot\mathbf{E}(\mathbf{r})dV$. To that
end, let us consider a spherical volume of the radius $r$,  such
that $l_0\ll r\ll \lambda$. At this scale the magnetic component
is not important and cannot be considered. In the sequel we will
work with dimensionless variable $r/l_0\rightarrow r$. The
electric field does not change at this scaling. A fractal mass
inside the volume is $\clM(r)\sim r^D$, where $0<D<3$. Therefore,
an average density of the metallic phase is of the order of
$r^{D-3}$. For the filtering mass inside the fractal, we consider
convolution of Eq. (\ref{fpe3}). We also, reasonably, suppose that
the random fractal composite is isotropic, \textit{i.e.}, the
(averaged) similarity exponents coincides along all directions.
This yields that one takes into account only radius $r$ and any
changes in the inclination and azimuth angle directions can be
neglected. Then, we have for the divergence
\be\label{qfsef2}  %
\nabla\cdot\mathbf{E}_1(\mathbf{r})=
\nabla_rE_{r,1}(r)\equiv\nabla_rE_1(r) \, . \ee   %
Now, filtering inside the fractal due to characteristic function
depends only on the radius $r$, and this yields the integrations
\footnote{In Cartesian's coordinates it corresponds to the
fractional Riemann-Liouville integral with the elementary
fractional volume \cite{SKM,tar2005}
$dV_D=\frac{|xyz|^{D/3-1}}{\Gamma^3(D/3)}dxdydz$. In the spherical
coordinates, which corresponds to the Reisz definition of the
fractional integral, the elementary fractional volume is
$dV_D=\frac{2^{3-D}\Gamma(3/2)}{\Gamma(D/2)}|\mathbf{r}|^{D-3}r^2dr\sin\theta
d\theta d\phi$.}
\bea\label{qsef3}   %
&\frac{1}{4\pi}\int\chi(\mathbf{r})\nabla_rE_1(r)dV =
\int_0^r\chi(r')[\nabla_rE_1(r')]{r'}^2dr' \nonumber \\
&=\int_0^r\sum_{r_j\in V_D} \delta(r'-r_j)G'(r')dr' \eea %
Here we define
%
$G(r)=r^2E_1(r)$  
and $G'(r)\equiv \frac{d}{d\,r}G$. One obtains the integration of
the electric field with the fractal density $\sum_{r_j\in V_D}
\delta(r'-r_j)$. The latter corresponds to the fractal volume
\be\label{qfse3a}  %
\mu(r)=r^D=\int_0^r\sum_{r_j\in V_D} \delta(r'-r_j){r'}^2dr'\,
. \ee  %
Following Ref. \cite{Ren} (Theorem $3.1$), we obtain
\bea\label{theorRen}  %
\int_0^rG'(r')d\mu(r')&\sim&\frac{1}{\Gamma(D-2)}
\int_0^r(r-r')^{D-3}G'(r')dr' \nonumber \\
&\equiv& {}_0I_r^{D-2}G'(r)\, . \eea   %
Therefore, we consider the integration in Eq. (\ref{qsef3}) as the
convolution integral with the averaged fractal density
$(r-r')^{D-3}$.

Now we estimate the integral
$\int_0^rE(r')\nabla_{r'}\chi(r'){r'}^2dr'$. The fractal dust
$V_{D}$ at the $N$th step of the construction consists of balls
$B_N$ with  the radius $\Delta_N$. For example, $\Delta_N\sim
l_0$. In the limiting case one obtains
$V_{D}=\lim_{N\to\infty}\bigcup B_N$ \cite{falconer}. The
characteristic function for every ball is $\chi(\Delta_N)=
\Theta(r-r_j)-\Theta(r-r_j-\Delta_N)$. Differentiation of the
characteristic function on the intervals $[r_j,r_j+\Delta_N]$
yields $ \nabla_r\chi(\Delta_N)
=\delta(r-r_j)-\delta(r-r_j-\Delta_N)$. Therefore, for any
interval $\Delta_N$ and at $r=r_j$, integration with the electric
field yields
$${}_rI_{r+\Delta_N}^1E(r)r^2\nabla_r\chi(r)
=E(r)r^2-E(r+\Delta_N)(r+\Delta_N)^2\, . $$   %
This expression is
not zero in the limit $\Delta_N\rightarrow 0$. Let $E(r_j)$ is the
electric field outside the ball $B_N$ and $E(r_j+\Delta_N)$
denotes the internal electric field. The relation between them,
due to Eq. (8.2) in Ref.~\cite{LLVIII} for polarization of a
dielectric ball, is
$$E(r_j+\Delta_N)=\frac{3\vep_d}{\vep_m(\omega)+2\vep_d}E(r_j)\, .
$$  
Therefore, the shift for the electric field in the limit
$\Delta_N\rightarrow 0$ is
\be\label{qsef4}  %
E(r_j)-E(r_j+0)=E(r_j)\frac{\vep_m(\omega)-
\vep_d}{\vep_m(\omega)+2\vep_d}\, . \ee    %
Finally, integration of the polarization charge term yields
\bea\label{qsef5} %
&\int_0^rE(r')\nabla_{r'}\chi(r')r'^2dr'\equiv
{}_0I_{r}^1E(r)r^2\nabla_r\chi(r) \nonumber  \\
&=\frac{\vep_m(\omega)-
\vep_d}{\vep_m(\omega)+2\vep_d}\sum_{r_j\in
V_D}\int_0^rE(r'){r'}^2\delta(r'-r_j)dr'\, . \eea  %
Again, one obtains the integration of the electric field with the
fractal density $\sum_{r_j\in V_D} \delta(r'-r_j)$. This
corresponds to the fractal volume (\ref{qfse3a}), and, hence, we
consider the integration in Eq. (\ref{qsef5}) as the convolution
integral of Eq. (\ref{theorRen})  with the averaged fractal
density $(r-r')^{D-3}$. Eventually, we obtain for the polarization
charge term in Eq. (\ref{qsef1_a})
\bea\label{qsef6}   %
&\int_0^rE(r')\nabla_{r'}\chi(r')r'^2dr'\sim  \nonumber \\
&-\frac{p(\omega)}{\Gamma(D-2)}\int_0^r(r-r')^{D-3}E(r'){r'}^2dr'
\nonumber  \\
&\equiv {}_0I_r^{D-2}[G(r)+\tilde{E}_{0}r^2]\,
. \eea   %
Here $\tilde{E}_0=\tilde{\mathbf{E}}\cdot\hat{\mathbf{r}}$ is a
projection of the external electric field on the radial direction
inside the chosen spherical volume of the radius $r$ and
\be\label{p_omega}  %
p(\omega)=\frac{\vep_d-\vep_m(\omega)}{\vep_m(\omega)+2\vep_d}
\, .  \ee  %

Taking all these arguments and results in Eqs. (\ref{theorRen})
and (\ref{qsef6}), one presents Eq. (\ref{qsef1_a}) in the
coarse-graining form
\bea\label{qsef7}   %
{}_0I_r^{D-2}G'(r)-q(\omega){}_0I_r^1G'(r)&-&
p(\omega){}_0I_r^{D-2}G(r) \nonumber  \\
&=& \frac{2p(\omega)\tilde{E}_0}{\Gamma(D+1)}r^D\, .\eea %
The Laplace transform can be applied to Eq. (\ref{qsef7}). Since
$G(r=0)=G'(r=0)=0$ (the electric field of fractal charge density
diverges slowly than $\frac{1}{r^2}$ \cite{bi2011,tar2005}), one
obtains for $\tlG(s)=\hclL[G(r)]$ due to Eq. (\ref{mt3}):
\be\label{qsef8}   %
\tlG(s)=\frac{2p(\omega)\tilde{E}_0}{\vep_ds^{3}}
\cdot\frac{1}{s-q(\omega)s^{D-2}-p(\omega)}\, .  \ee  %
Note that the second term in Eq. (\ref{qsef7}) is $q(\omega)G(r)$.

Before arriving at the main result, let us consider the limiting
cases. For $\vep_m(\omega)=\vep_d$ one obtains $q(\omega)=\infty$
and $p(\omega)=0$. This yields $E_1(r)=0$, and the solution for
the electric field is $E(r)=\tilde{E}_0\equiv\tilde{E}_r$. Another
limiting case is $|\vep_m(\omega)|\rightarrow\infty$. In this case
$q(\omega)=0$ and $p(\omega)=-1$, thus $E_1\sim \tilde{E}_0$.
Important result here is that permittivity of the mixture is
approximately $\vep_d$ that corresponds to the well known result
in Ref.~\cite{LLVIII} [see Eq. (9.7)].

\section{Surface plasmon resonance}
The most interesting case is the SPR, when
$Re[\vep(\omega)]=-2\vep_d$. This also known as the Fr\"ohlich
resonance \cite{Frohlich,bohren}. The permittivity of the fractal
metallic nanostructure at the resonance condition is a complex
value $\vep_m(\omega)=\vep_1+i\vep_2$, where $\vep_2/\vep_1\ll 1$
and is described by classical Drude formula\footnote{The
dependence of the permittivity of the fractal metallic
nonostructure on the frequency of the external electric field is
described by Drude formula (see \textit{e.g.}
\cite{LLVIII,bohren}). For a spherical volume it reads
$\vep_m(\omega)=\ep_0-\frac{\omega_p^2}{\omega(\omega+i\gamma)}$,
where $\omega_p$ is a so-called plasma frequency, $\ep_0$ is a
high-frequency lattice dielectric constant, while the attenuation
coefficient $\gamma$ is small in comparison with the resonant
frequency. Therefore, $\vep_m(\omega)=\vep_1+i\vep_2$, where
$\vep_1={\rm Re}[\vep_m(\omega)]=\ep_0-\omega_p^2/\omega^2$ and
$\vep_2={\rm Im}[\vep_m(\omega)]=\gamma\omega_p^2/\omega^3$.} . At
a small detuning from  the resonance, when
$Re[\vep(\omega)]=-2\vep_d+\vep_2$ that corresponds to the width
of the resonance in the frequency domain, $p(\omega)$ reaches the
maximal values, that yields
\bea\label{spr1}  %
p(\omega)=-1
+\frac{3\vep_d}{2\vep_2}(1-i)\approx\frac{3\vep_d}{2\vep_2}(1-i)\,
, \nonumber   \\
q(\omega)=\frac{1}{3}-\frac{\vep_2}{9\vep_d}(1-i)
\approx \frac{1}{3}\, . \eea  %
These expressions are inserted in Eq. (\ref{qsef8}). Before
carrying out the inverse Laplace transform $\hclL^{-1}[\tlG(s)]$,
it is reasonable to simplify the second denominator. We recast Eq.
(\ref{qsef8}) in the form
\be\label{spr2}  %
\tlG(s)=2p(\omega)\tilde{E}_0\sum_{k=0}^{\infty}
\frac{[q(\omega)s^{D-2}+p(\omega)]^k}{s^{k+4}}\, .  \ee  %
We take into account that for the scale $r\gg 1$, the Laplace
parameter is small $s\ll 1$, and the binomial becomes
approximately a monomial. For $D<2$, the first term can be left,
while for $D\geq 2$ the second term $p(\omega)$ is dominant. To
keep the first term for $D<2$, we have $(1/s)\sim r\gg
(3\vep_d/2\vep_2)^{\frac{1}{2-D}}$. But this condition violates
another restriction for $r\ll\lambda/l_0$, and $p(\omega)$ is the
most important term in the binomial, and cannot be omitted. The
Laplace inversion of Eq. (\ref{spr2}) can be performed using an
expression for the Mittag-Leffler function \cite{BE}
\bea\label{spr3}  %
\clE_{(\nu,\beta)}(zr^\nu)= \frac{r^{1-\beta}}{2\pi
i}\int_{\clC}\frac{s^{\nu-\beta}e^{sr}}{s^{\nu}-z}ds=
\nonumber \\
\frac{r^{1-\beta}}{2\pi}\int_{\clC}e^{sr}\sum_{k=0}^{\infty}
\frac{z^k}{s^{\nu k+\beta}}ds
=\sum_{k=0}^{\infty}\frac{[zr^{\nu}]^k}{\Gamma(\nu k+\beta)}\, , \eea %
where $\clC$ is a suitable contour of integration, starting and
finishing at $-\infty$ and $\nu,\beta>0$.  Comparing Eqs.
(\ref{spr2}) and (\ref{spr3}), $\beta=4$ and $\nu=1$, one obtains
for the electric field
\be\label{spr4}    %
E_1(r)=2p(\omega)\tilde{E}_0r\clE_{(1,4)}[p(\omega)r]\, . \ee  %
Since the argument of the Mittag-Leffler function is large, its
asymptotic behavior is \cite{BE}
\be\label{spr5}  %
E_1(r)\sim \frac{2\tilde{E}_0r}{p^2(\omega)}e^{p(\omega)r}\propto
\frac{E_0}{\vep_d}\exp\left[r\frac{3\vep_d}{2\vep_2}(1-i)\right]\,
.\ee   %
Eventually, we arrived at the exponential (geometrical)
enhancement and giant oscillations of the respond electric field
due to the fractal geometry of the metal-dielectric composite.
This result is valid for $r\geq 1$. At the same condition one can
also choose the SPR at ${\rm
Re}\,\vep_m(\omega)=-2\vep_d-\eta\vep_2$, where $\eta$ is defined
from the maximum enhancement of the electric field. This yields
$q(\omega)\approx 1/3$ and $p(\omega)\approx\vep_d/\vep_2$. Thus
the first term in the denominator in Eq. (\ref{qsef8}) can be
neglected and the latter can be recast in the form
\be\label{spr6} %
\tlG(s)=-\frac{2p(\omega)\tilde{E}_0}{q(\omega)}\sum_{k=0}^{\infty}
\frac{[-p(\omega)/q(\omega)]^k}{s^{(k+1)(D-2)+3}}\, .  \ee  %
This yields the following expression for the electric field
\be\label{spr7}  %
E_1(r)=\frac{2p(\omega)\tilde{E}_0r^{D-2}}{q(\omega)}
\clE_{(D-2,3)}[-p(\omega)r^{D-2}/q(\omega)]\, .    \ee  %
If $\eta$ is chosen such that ${\rm
arg}[\frac{-p(\omega)}{q(omega)}]<\frac{(D-2)\pi}{2}$, the
asymptotic behavior of the Mittag-Leffler function is exponential
\cite{BE} that leads to the giant geometrical enhancement
$E_1(r)\sim\exp \left[\left(
\frac{3\vep_d}{\vep_2}\right)^{\frac{1}{D-2}}r\right]$ that,
obviously, takes place at the condition $D>2$.

When the resonance is exact and the detuning from the resonance is
zero, then ${\rm Re}[p(\omega)]=-1$. The arguments of the
Mittag-Leffler functions in both equations (\ref{spr4}) and
(\ref{spr7}) do not correspond to the exponential asymptotic
behavior. In this case the geometrical enhancement of the electric
field is absent, and the linear (``classical'') enhancement
\cite{bohren} of the electric field takes place due to the
imaginary part of the pre-factor: ${\rm Im\,}p(\omega)\sim
\frac{\vep_d}{\vep_2}\gg 1$.

\section{Discussion on the geometrical enhancement of the
electric field}

The obtained expressions in Eqs. (\ref{spr2}) and (\ref{spr4})
(together with developing the coarse grained Maxwell equation in
the form of Eqs. (\ref{qsef7}) and (\ref{qsef8})) contain the main
physical result. As seen, the resulting (enhanced) electric field
depends on the parameter $p(\omega)$, which is the pre-factor and
the argument of the Mittag-Leffler function (when $r\sim 1$) in
Eqs. (\ref{spr4}) and (\ref{spr7}), and defined in Eq.
(\ref{p_omega}). This parameter is absorbtion efficiency in the
electrostatic (quasi-static) approximation \cite{bohren}. In our
case, it describes polarization, and it is obtained in Eq.
(\ref{qsef5}) under evaluation of fractal boundary conditions. One
should recognize that the linear (classical \cite{bohren})
enhancement of the electric field is always takes place, due to
Eq. (\ref{qsef5}) that corresponds to the enhancement of the
electric field by a small particle and that is reflected by the
pre-factor in Eq. (\ref{spr2}). For a fractal small composite of
many particles the situation differs essentially. Here the
geometrical enhancement of the electric field is due to the
Mittag-Leffler function and it depends on the argument of the
complex value of $p(\omega)$.  The exponential (geometric)
enhancement takes place when $|p(\omega)|\gg 1$ and ${\rm Re}
p(\omega)>0$ as in Eq. (\ref{spr4}), or ${\rm Re}[-p(\omega)]>0$
as in Eq. (\ref{spr7}). These conditions are fulfilled for those
frequencies $\omega$ that are in the vicinity of the SPR: ${\rm
Re}p(\omega)=-2\vep_d+\Delta$, where $\Delta\sim\vep_2\ll \vep_d$.

We have to admit that the obtained expressions in the exponential
forms are the upper bound of the electric field enhancement. The
enhanced $E_1$ does not exceed of the order of $10^8$(v/cm),
otherwise the nonlinear effects become important that violates the
linear quasi-static consideration. This restriction yields
$r(\vep_d/2\vep_2)\leq 20$, which is a reasonable value for
experimental realizations. Therefore, we have the light wavelength
$\lambda\sim 10^{-4}$cm, fractal inhomogeneity size $l_0\sim
(10^{-6}\div 10^{-5})$cm, and $\vep_d/\vep_2\sim\omega\tau\sim
5\div 15$, where $\omega$ is the optical frequency, while $\tau$
is the relaxation time. The latter value determines $l_0$, as
well, which was introduced above as a minimum self-similarity
size. Note that $\tau\leq \tau_s\equiv\frac{l_0}{v_F}$, where
$\tau_s$ is the surface relaxation time, while $v_F$ is the
velocity on the Fermi surface (for free electrons). Therefore,
from the condition $\omega\tau\gg 1$, one obtains $l_0\gg
\frac{v_f}{\omega}\sim 10^{-7}$cm.

\subsection{Geometrical enhancement out of the SPR}

Equation (\ref{spr4}) describes geometrical enhancement of the
electric field out of the SPR, as well. For those frequencies
$\omega$ that $p(\omega)> 1$, the argument of the Mittag-Leffler
function is large, when $1\ll r \ll \frac{\lambda}{l_0}$ that
leads (roughly) to the exponential $E_1(r)\sim
\tilde{E}_0\exp[p(\omega) r]$. For example, when
$\vep_m(\omega)=-\vep_d$, the small imaginary part of
$\vep_m(\omega)$ is not important, and the condition for the large
asymptotics of the Mittag-Leffler function is fulfilled. One
obtains  $p(\omega)\approx 2$ and $E_1(r)\sim \tilde{E}_0\exp[2
r]$ that yields an enhancement of the order of $10^4$ for $r>5$.

\section{Conclusion}
We developed an analytical approach, for description of the wave
propagation in metal-dielectric nanostructures in the
\textit{quasi-static} limit. The method is based on fractional
calculus and permits to obtain an analytical expressions for the
electric field enhancement. This approach establishes a link
between fractional geometry of the nanostructure and fractional
integro-differentiation. An essential (geometrical) enhancement of
the electric field is obtained for the surface plasmon resonance
at a certain relation between permittivities of the host and
fractal metallic nanostructure, when ${\rm
Re}\vep_m(\omega)=-2\vep_d$ with a suitable detuning.

Important part of the analysis is developing convolution integrals
that makes it possible to treat the fractal structure. The initial
Maxwell equation (\ref{qsef1}) is local, since $l_0/\lambda\ll 1$
and space heterogeneity is accounted locally by virtue of the
characteristic function. An accurate treatment of the fractal
boundaries and recasting the Maxwell equation in the form of the
convolution integrals by accounting fractal properties of the
composite, eventually, leads to the coarse graining equations,
which already take into account the space heterogeneity and
nonlocal nature of the electric field and polarized dipole charges
inside the composite. Therefore, the heterogeneity, caused by the
fractal geometry, is reflected by the convolution of the averaged
fractal density and the electric field, according
Eqs.(\ref{theorRen}) and (\ref{qsef6}). It is necessary to admit
that this transform from ``local'' quasi-electrostatics to the
nonlocal, which takes into account space dispersion of
permittivity $\vep(r)$ is mathematically justified and rigorous
enough \cite{Ren}. The obtained convolution integrals are averaged
values, since the fractal density $r^{D-3}$ is the averaged
characteristics of the fractal structure, and, eventually, it
determines the space dispersion of the permittivity $\vep(r)$ of
the mixture.

Summarizing, we have to admit that observation of macroscopic
Maxwell's equations is related to averaging of microscopic
equations \cite{LLVIII}. This procedure for fractal composite
media is not well defined so far, since, according fractal's
definition, averaging over any finite volume depends on the size
of this volume itself \cite{falconer}. The main idea to overcome
this obstacle is to refuse the local properties of equations and
obtain nonlocal coarse graining Maxwell's equations, which are
already averaged.

\acknowledgments This research was supported in part by the Israel
Science Foundation (ISF) and by the US-Israel Binational Science
Foundation (BSF).

\end{document}